\documentstyle[11pt,aaspp4,epsfig]{article}
\slugcomment{Published in ApJ Letters (1998, 492, L53)}

\begin{document}

\title{EVIDENCE FOR FRAME-DRAGGING AROUND SPINNING BLACK HOLES IN X-RAY BINARIES}
\author{Wei Cui\altaffilmark{1}, S. N. Zhang\altaffilmark{2,3}, and
Wan Chen\altaffilmark{4,5}}

\altaffiltext{1}{Center for Space Research, Massachusetts Institute of 
Technology, Cambridge, MA 02139}

\altaffiltext{2}{ES-84, NASA/Marshall Space Flight Center, Huntsville,
AL 35812}

\altaffiltext{3}{Universities Space Research Association}

\altaffiltext{4}{Department of Astronomy, University of Maryland, College
 Park, MD 20742}
\altaffiltext{5}{NASA/Goddard Space Flight Center, Code 661, Greenbelt,
 MD 20771}

\begin{abstract}

In the context of black hole spin in X-ray binaries, we propose that certain 
type of quasi-period oscillations (QPOs) observed in the light curves of black
hole binaries (BHBs) are produced by X-ray modulation at the precession 
frequency of accretion disks, due to relativistic dragging of inertial frames 
around spinning black holes. These QPOs tend to be 
relatively stable in their centroid frequencies. They have been observed in 
the frequency range of a few to a few hundred Hz for several black holes with 
dynamically determined masses. By comparing the computed disk precession 
frequency with that of the observed QPO, we can derive the black hole angular
momentum, given its mass. When applying this model to GRO J1655-40, 
GRS 1915+105, 
Cyg X-1, and GS 1124-68, we found that the black holes in GRO J1655-40
and GRS 1915+105, the only known BHBs that occasionally produce superluminal
radio jets, spin at a rate close to the maximum limit, while Cyg X-1 and
GS 1124-68, typical (persistent and transient) BHBs, contain only moderately 
rotating ones. Extending the model to the general population of 
black hole candidates, the fact that only low-frequency QPOs have been detected
is consistent with the presence of only slowly spinning black holes in these
systems. Our results are in good agreement with those derived from spectral
data, thus strongly support the classification scheme that we proposed 
previously for BHBs.

\end{abstract}


\section{Introduction}

It has long been argued that a spinning black hole may provide an ideal
laboratory for testing the theory of general relativity (GR). Recently, 
Zhang, Cui, \& Chen (1997; Paper I hereafter) have successfully derived
the angular momentum of accreting black holes (BHs) in several Galactic X-ray 
binaries, by carefully modeling the soft disk component in the observed 
X-ray spectra for high luminosity states. They found that superluminal 
jet sources contain BHs spinning close to the extreme theoretical limit 
while most others contain only slowly rotating ones. For BHBs, 
the relativistic dragging of inertial frames around a spinning BH should 
cause the accretion disk to precess, if it does not coalign with the BH 
equatorial plane, thus may produce direct observable effects.

Recently, Stella \& Vietri (1997) proposed that for low-mass X-ray binaries 
that contain an accreting neutron star (LMXBNs) the observed QPOs in the 
range of tens of Hz are the manifestation of disk precession due to the 
``frame dragging'' (FD) effect. They were able to account for such QPOs 
for several atoll and Z sources, with the help of recent discoveries of kHz 
QPOs in LMXBNs and progress in theoretical interpretation of such phenomena. 
Inspired by their work, we quickly realized that the same may also apply to 
certain type of QPOs that seem to be unique to BHBs.

Though not as commonly seen as in LMXBNs, QPOs have been observed in several
BHBs over a frequency range from mHz to roughly ten Hz (review by van der 
Klis 1995, and references therein). With much improved timing resolution of 
the RXTE instrumentation (Bradt, Rothschild, \& Swank 1993), the upper range 
has recently been pushed up to hundreds of Hz (Remillard et al. 1997). The QPO models
that invoke magnetic effects for LMXBNs probably do not apply to BHBs, because
of the less importance of the magnetic field, which is maintained by the 
accretion disk, in such systems. Consequently, accretion disks in BHBs can 
probably extend all the way (or very close) to the last stable orbit, under 
certain conditions. Therefore, we expect GR effects to be stronger in BHBs 
than in LMXBNs.

One type of QPO in BHBs is of particular interest. The QPO was first reliably
established in GRS 1915+105, a superluminal jet source, with the detection of
now famous 67 Hz QPO (Morgan, Remillard, \& Greiner 1997). One unique 
characteristic of this QPO is the constancy of its centroid frequency (which
moves around by only a few Hz). The QPO is clearly a transient event, and is 
only present in certain spectral states. The origin of such a QPO is still
unknown. Shortly after this discovery, a similar QPO was detected at $\sim$300
Hz for GRO J1655-40, another superluminal jet source (Remillard et al. 1997). 
This 
led to the speculation that such phenomenon may be common for BHBs and may be 
related to some fundamental properties of these sources. In retrospect, we may 
have seen similar QPOs in other BHBs before as well, such as Cyg X-1 (Cui et 
al. 1997) and GS 1124-68 (Belloni et al. 1997), although the QPOs are not as 
stable (the centroid frequency can vary by up to a factor of two). A more 
stable QPO was also seen at 6 Hz for GX 339-4, a BH candidate (BHC), in the 
very high state (Miyamoto et al. 1991). It seems natural to ask if all of 
these QPOs are indeed similar in origin and what are the physical processes 
that produce them. In this letter, we propose that they are the result of
disk precession due to the FD effect. When applying this model to several 
known BHs with dynamically determined masses, we can derive their angular 
momenta. 
We show that the results are consistent with what we derived from spectral 
data in Paper~I, thus providing further support to the classification scheme 
that we proposed for BHBs. 

\section{Accretion Disk and Frame Dragging Effect}

As in Paper~I, we assume a geometrically thin, optically thick accretion disk 
(Novikov \& Thorne 1973; Shakura \& Sunyaev 1973), which now may be tilted 
with respect to the equatorial plane of a spinning BH. Further more,
we assume that in the high luminosity state (i.e., {\it not} the advection
dominated state; e.g., Narayan \& Yi 1995) the inner disk edge extends to the 
last stable orbit. Of course, with a central X-ray source, radiation pressure
may have effects on the inner disk structure. We will delay the discussion 
of such effects until the next section.

The radius of the last (marginally) stable orbit of a test particle is a
function of the BH angular momentum (Bardeen, Press, \& Teukolsky 1972),  
\begin{equation} 
r_{\rm last} = r_{\rm g} \{3+A_{2}\pm[(3-A_{1})(3+A_{1}+2A_{2})]^{1/2}\},
\end{equation} 
where
$A_{1}=1+(1-a^{2}_{\ast})^{1/3}[(1+a_{\ast})^{1/3}+(1-a_{\ast})^{1/3}]$,
$A_{2}=(3a^{2}_{\ast}+A_{1}^{2})^{1/2}$; $a_{\ast} = a/r_{\rm g}$, with
$a=J/Mc$ ($J$ is the BH angular momentum and $M$ the mass) and 
$r_{\rm g}=GM/c^2$; the lower and upper signs are for prograde orbits (i.e., 
in the same sense as the BH spin, i.e., $a_{\ast}>0$) and retrograde 
orbits ($a_{\ast}<0$), respectively. In the presence of an accretion disk, 
$a_{\ast}$ takes values in the range from -0.998--+0.998 (Thorne 1974), 
with $\pm$1 being the absolute theoretical limits. The event horizon of a 
Kerr BH is at
$r_{\rm h}=r_{\rm g}+(r_{\rm g}^{2}-a^{2})^{1/2} = r_{\rm g} [ 1+ (1-a^2_{\ast})^{1/2}]$,
so the disk may extend all the way to the horizon, 
$r_{\rm last}(a_\ast$=+1)=$r_{\rm g}$, or be expelled to 
$r_{\rm last}(a_{\ast}$=--1)=9$r_{\rm g}$, as compared to the canonical 
Schwarzschild case, $r_{\rm last}(a_{\ast}$=0)=6$r_{\rm g}$. 

As discussed in Paper~I, X-ray emission from the disk is from the hot 
innermost region. The X-ray spectrum from such a disk can be
described as a ``diluted'' blackbody, but with a varying 
temperature as a function of radius. The effective temperature of the disk 
peaks at an annulus slightly beyond $r_{\rm last}$ at
$r_{\rm peak} = r_{\rm last}/\eta$, where $\eta$ varies only slowly from
0.62 to 0.76 as $a_{\ast}$ goes from -1 to +1. {\it That is where most of the
disk X-ray emission comes from.}

Due to the FD effect, the tilted orbital plane of a test particle must precess
around the same axis and in the same direction as the BH spin (Lense \& 
Thirring 1918; Wilkins 1972). For simplicity, we specialize to only circular 
orbits, which are most relevant to accretion processes in X-ray binaries. 
First, we define a node as a point where a non-equatorial orbit intersects the 
equatorial plane. Then, the nodal precession frequency can be expressed as
\begin{equation}
\nu_{FD} = \nu_{orb} \frac{\Delta \Omega}{2\pi}, 
\end{equation}
where $\nu_{orb}$ is the orbital 
frequency, and $\Delta \Omega$ is an angle by which the nodes of a circular 
orbit are dragged per revolution. In the weak field limit, Lense and Thirring 
(1972) have derived 
$\Delta \Omega / 2\pi = 2(\left|a\right|/c) (G M/r^3)^{1/2}$, 
where $r$ is the 
radius of test particle orbit. In this case, the orbital frequency is simply
that of Keplerian motion, i.e., $\nu_{orb} = (1/2\pi) (G M/r^3)^{1/2}$. 
Substituting these into Eq. (2), we derive the Lense-Thirring (LT) 
precession frequency 
\begin{equation}
\nu_{LT} = 6.45\times 10^4 \left|a_{\ast}\right| \left(\frac{M}{M_{\odot}}\right)^{-1} \left(\frac{r}{r_g}\right)^{-3} Hz.
\end{equation}

For BHBs that contain extremely spinning BHs ($a_{\ast} \simeq +1$), the
accretion disk can extend very close to the event horizon 
(at $r \simeq r_{\rm g}$), where weak field approximation clearly breaks down. 
The exact problem has been solved analytically, by Wilkins (1972), to derive
the allowed ranges for constants of motion, $E$, $\Phi$, and $Q$, which are
respectively the energy, the component of angular momentum along the BH spin
axis, and a non-negative quantity related to the $\theta$ velocity (with 
$Q=0$ specifying equatorial orbits). The requirement for a stable circular 
orbit provides two equations that allow the expression of $E$ and $\Phi$ in 
terms of $Q$ (and of course $r$ and $a_{\ast}$) (cf. Wilkins 1972). Therefore, 
unlike the weak-field limit, for a given $a_{\ast}$ it is no longer true in 
general that $\Delta \Omega$ depends only on the orbital radius. 

The orbital frequency of a test particle around a spinning BH also deviates 
from that of Keplerian motion at small radii. Bardeen et al. (1972) have
shown that the frequency is given by 
\begin{equation}
\nu_{orb} = 3.22\times 10^4 \left(\frac{M}{M_{\odot}}\right)^{-1} \left[\left(\frac{r}{r_g}\right)^{3/2} + a_{\ast} \right]^{-1} Hz.
\end{equation} 
With corrections to both terms in Eq. (2) in the case of strong field, we can 
compute $\nu_{FD}$ as a function of $r$, given $a_{\ast}$, $Q$, and
BH mass. As an example, we plot the results in Fig.~1, for cases where 
$a_{\ast}=+0.5, +0.95, +1$, and $Q=0.01$ (i.e., slightly off the equatorial
plane). For comparison, we have also plotted the weak field approximation. As 
expected, strong-field effects are only important for 
extremely spinning BHs. Since the bulk of disk emission is produced at
$r_{\rm peak}$, we have derived the precession frequency at this radius, as 
shown in Fig.~2. It is interesting to note that the frequency is always below 
$\sim$8 Hz (derived for a $3 M_{\odot}$ BH) for systems with retrograde disks,
and at low frequencies there are two solutions to the BH angular momentum for 
a given mass.

\section{Disk Precession and QPOs}

We now propose that the ``stable'' QPOs observed in BHBs are simply 
X-ray modulation at the disk precession frequency. First, we apply this
hypothesis to micro-quasars, GRO~J1655-40 and GRS~1915+105, since we know 
from Paper~I that both may contain nearly maximal rotating BHs. The BH mass 
of GRO~J1655-40 
was determined to a high accuracy, $7.02\pm 0.22 M_{\odot}$ (Orosz \& Bailyn 
1997). If the observed QPO frequency (300 Hz) is indeed the disk precession 
frequency at $r_{\rm peak}$, we derive $a_\ast = +0.95$, in 
excellent agreement with the most probably value from Paper~I. The same can 
be applied to GRS~1915+105, although the BH mass is not reliably determined 
in this case. The 67 Hz QPO implies $a_\ast \simeq +0.65$ for a 3$M_{\odot}$ 
BH or +0.95 for a 30$M_{\odot}$ one. Compared with the spectral results
from Paper~I, a self-consistent solution for the BH mass should be around
30 $M_{\odot}$. While the same value of $a_\ast$ for both sources may purely 
be a numerical coincidence, it strongly hints at the presence of an rapidly 
spinning BH in GRS~1915+105 and extraordinary similarities between the two. 

Then, we examine more typical BHBs. A strong QPO was observed for GS 1124-68 
during the very high state (VHS) and a transition from the high to low state 
(Belloni et al. 1997). Within each observation, the QPO frequency remained 
quite stable, but it varied in the range 5--8 Hz between observations. The 
QPO observed during the transition (at 6.7 Hz) appears to be of the same 
origin. In VHS, the mass accretion rate is thought to be near the Eddington
limit (van der Klis 1995), thus radiation pressure may play an important
role in shaping up the inner disk structure. Depending on the exact details of
the physical processes involved, VHS might not be a long-term stable state 
as far as the accretion disk is concerned. The inner edge of the optically 
thick disk may sometimes be pushed outward (Misra \& Melia 
1997), consequently the disk precession frequency 
becomes smaller. We would then expect an anti-correlation between the QPO 
frequency and X-ray luminosity; there seems to be a hint of such relationship
for GS 1124-68 (Belloni et al. 1997). From
Eq. (3) it is clear that disk precession frequency is a sensitive function 
of position ($r$) --- it only takes a variation of $\sim 17$\% in radius to 
account for the frequency variation observed. The BH mass of the 
source was measured to be $\sim 6.3 M_{\odot}$ (Orosz et al. 1996), so a QPO 
frequency of 8 Hz would imply $a_\ast = +0.35$, which is moderate
compared to that of superluminal jet sources. A similar QPO was also observed 
for GX 339-4 also in VHS (Miyamoto et al. 1991). The measured QPO frequency 
clusters around 6 Hz. It should be noted, however, that evidence for 
the source being a BHB is at best circumstantial --- no dynamical BH mass 
measurement is available. Nevertheless, the source has long been considered a 
BHC, because of the similarities to Cyg X-1 in its X-ray properties.

Recently, a QPO has been detected in Cyg X-1 {\it only} during the transitions 
between its hard and soft states (Cui et al. 1997), perhaps similar to that
in GS~1124-68 observed during the transition. The QPO frequency varied 
in the range 4 -- 9 Hz, showing strong correlation with X-ray spectral shape.
This is consistent with the fact that the evolution of the accretion disk was 
not completed yet during these episodes; consequently the innermost disk may 
still be in the process of settling down to its final stable configuration. 
Given the BH mass $10 M_{\odot}$ (Herrero et al. 1995), a QPO frequency of
9 Hz indicates $a_\ast = +0.48$, which is lower than what we 
inferred (+0.75) from the spectral data for the soft state. This is,
nonetheless, an encouraging result considering the uncertainties (on the BH 
mass, the inclination angle, and so on) involved in the calculations. If the 
spectral state transitions are indeed caused by flip-flops of the disk, as 
we suggested in Paper~I, we might see a QPO at $\sim$2 Hz in the hard state 
(for $a_\ast = -0.48$). Of course, the detection
is not guaranteed since mechanisms for producing X-ray modulation might only
be present under certain conditions. For instance, no apparent QPOs have been
detected in the soft state (Cui et al. 1997). 

Only low-frequency QPOs (at a few to tens of mHz) have been detected for
a few other BHCs (van der Klis 1995). This seems to be consistent with most 
BHBs containing only slowly rotating BHs (see Paper~I), although it is not 
clear if any of these QPOs are due to disk precession.

\section{Discussion}

We summarize the results for known BHBs in Table~1. Although the results are 
computed for a specific $Q$ value ($Q=0.01$), they are actually quite 
insensitive to $Q$. For example, we have calculated the angular momentum for 
GRO~J1655-40 as a function of allowed $Q$ values (0 -- 2) for stable circular 
orbits. Over the entire range, the result varies only by roughly 1\%. 

Our results suggest that the FD effect is quite significant in BHBs and is 
manifested in the presence of relatively stable QPOs in these systems. We 
emphasize that the strength of the disk precession model lies in the fact
that it can account for certain type of QPOs in {\it all} BHBs. Another
model invoking trapped $g$-mode oscillations near the inner edge of the 
disk can also quantitatively account for the 300 Hz QPO for GRO J1655-40 
and 67 Hz QPO for GRS 1915+105 (Nowak et al. 1997; Paper~I), but not
those in other BHBs, such as Cyg X-1 and GS 1124-68 (cf. Perez et al. 1997). 
Therefore, we consider the disk precession model being more natural for BHBs. 
The model provides a direct way of measuring BH angular momentum from the 
observed QPO, once the mass is determined, independent
of measurements based on spectral information. It is the consistency between 
the two types of measurements (see Table~1) that gives us confidence in the 
model.

Inferring BH angular momentum directly from QPOs avoids the need for 
information on binary
orbital inclination, thus does not suffer from the large uncertainty associated
with inclination angles, which is built into spectroscopic 
measurements (Paper~I). Compared with Paper~I, a slightly lower value for 
$a_{\ast}$, from the 67 Hz QPO for GRS 1915+105, might suggest an actual 
inclination angle slightly larger than 70\arcdeg\ for the system, while a 
higher $a_{\ast}$ for GS~1124-68 might be due partly to a smaller inclination 
(than that adopted in Paper~I).

The results show that the BHs in superluminal jet sources spin near the 
maximum limit, while those in ``normal'' BHBs only moderately or slowly 
rotate. This supports the idea that jet 
formation may be closely related to the BH spin. Some models suggest that
jets are formed by the ejection of matter from the inner disk region 
and are collimated by the magnetic field maintained by the disk (e.g. 
Blandford \& Payne 1982; Shu et al. 1995; Ustyugova et al. 1995; Kudoh \&
Shibata 1997). If so, the 
disk precession model would naturally predict the precession of jets at the
same frequency in these systems. Future observations will shed light on this 
issue. 

Why do most BHBs only contain slowly or moderately spinning BHs? There are 
two important 
spin-down mechanisms: accretion from retrograde disks onto BHs (Moderski \& 
Sikora 1996) and the Blandford-Znajek extraction of rotational energy of 
spinning BHs (Blandford \& Znajek 1977; Moderski, Sikora, \& Lasota 1997). 
It has been shown that it is much easier to spin down a BH by accreting from 
a retrograde disk than to spin it up from a prograde disk (Thorne 1974; 
Moderski \& Sikora 1996). For instance, a BH must accrete about 20\% 
of its initial mass from a retrograde disk to decelerate from maximum to
zero spin, while it takes roughly 180\% of its current mass 
from a prograde disk to accelerate it back up to maximum spin. This 
process alone can cause a uniform initial distribution of BH angular momentum 
to evolve
into an extremely non-uniform one with most systems being relatively slowly 
rotating; the Blandford-Znajek process would only speed up such evolution, 
despite being relatively inefficient (Moderski, Sikora, \& Lasota 1997). 
In reality, mass accretion is probably inadequate in spinning up those 
born slow rotators to the extreme limit, due to such factors as low average 
accretion rate, disk flip-flop, or the lack of available mass from companion 
stars. This implies that for systems like GRO~J1655-40 and GRS~1915+105 BHs 
are likely formed with high angular momentum.

One key element to the disk precession model is the requirement of misalignment
of the disk with the equatorial plane of a spinning BH. Such misalignment 
could be the result of the Pringle instability (Pringle 1996; also see Stella 
\& Vietri 1997 for detailed discussion). However, the Bardeen-Petterson effect
(Bardeen \& Petterson 1975) should lead to the equatorial configuration for 
the innermost disk, thus the precession might cease for the X-ray emitting 
region (although the time scale for it to happen seems to be quite long; see
Scheuer \& Feiler 1996). This might be the reason why the QPOs of interest 
only occur during
transitional or unstable periods, when the inner disk region experiences
significant changes, and why they only last for a limited time. Moreover, the 
disk precession only provides a natural frequency for QPOs, and physical 
mechanisms are still needed
to produce X-ray modulations. At present we have no definitive knowledge about
the mechanisms, but the modulation could be caused by any kind of asymmetry or
inhomogeneity in a precessing disk, as well as by the varying projected area 
of the disk ring and/or perhaps an occulted (by the disk) central hard X-ray 
emitting region (cf. Stella \& Vietri 1997). 

Finally, it is interesting to compare BHBs with atoll sources that first drew 
Stella and Vietri's attention. For atoll sources, the magnetic field is 
thought to be very weak (e.g., van der Klis 1995), it may hardly affect the 
dynamics of accretion flows. Consequently, these system should
resemble BHBs in their X-ray properties including QPOs. Indeed, many 
similarities between the two have been observed, such as hard 
power-law tails in the X-ray spectra (e.g., Zhang et al. 1996), 
which used to be considered one of the defining signatures for BHCs. 
It is therefore of no surprise that the disk precession model works well in 
both types of systems. However, no QPOs at the Keplerian frequency of the
inner disk edge (appearing as kHz QPOs in LMXBNs) are present in BHBs. This 
might be related to the lack of magnetic field in BHBs, if the interaction 
between magnetosphere and accretion disk is responsible for introducing 
X-ray modulation. The situation becomes more complicated for LMXBNs with 
relatively strong field (such as Z sources), since in those cases precise
knowledge about the field strength is required to determine the location of 
the inner disk boundary. Recent RXTE discoveries of kHz QPOs in Z sources 
seem to provide such information (Stellar \& Vietri 1997). 

\acknowledgments
We thank Luigi Stella and Mario Vietri for their inspiring work that triggered
this investigation, and the referee for prompt and helpful comments. 
Cui acknowledges support by NASA through Contract NAS5-30612. SNZ is 
partially supported by NASA Grants NAG5-3681, NAG5-4411, and NAG5-4423.

\clearpage

\begin{deluxetable}{lcccc}
\tablecolumns{5}
\tablewidth{0pc}
\tablecaption{Inferred Black Hole Angular Momentum\tablenotemark{*}}
\tablehead{
\colhead{Source}&\colhead{Mass}&\colhead{QPO Frequency}&\colhead{$a_{\ast}$} &\colhead{References}\\ 
 & ($M_{\odot}$) & (Hz) & & }
\startdata
GRO J1655-40 & $7$  & $300$ & $+0.95$ ($+0.93$)& 1, 2\\
GRS 1915+105 & $30$ & $67$ & $+0.95$ ($\sim +1$)& 3, 4\\
GS 1124-68 & $6.3$  & $8$ & $+0.35$ ($-0.04$)& 5, 6 \\
Cyg X-1 & $10$      & $9$ & $+0.48$ ($+0.75$)& 7, 8 \\
\tablenotetext{*}{The numbers in parentheses are the results from Paper~I.}
\tablenotetext{1}{Orosz \& Bailyn 1997.} 
\tablenotetext{2}{Remillard et al. 1997.}
\tablenotetext{3}{Paper~I.} 
\tablenotetext{4}{Morgan, Remillard, \& Greiner 1997.}
\tablenotetext{5}{Orosz et al. 1996.} 
\tablenotetext{6}{Miyamoto et al. 1994.}
\tablenotetext{7}{Herrero et al. 1995.} 
\tablenotetext{8}{Cui et al. 1997.}
\enddata
\end{deluxetable}

\clearpage
\begin{figure}
\psfig{figure=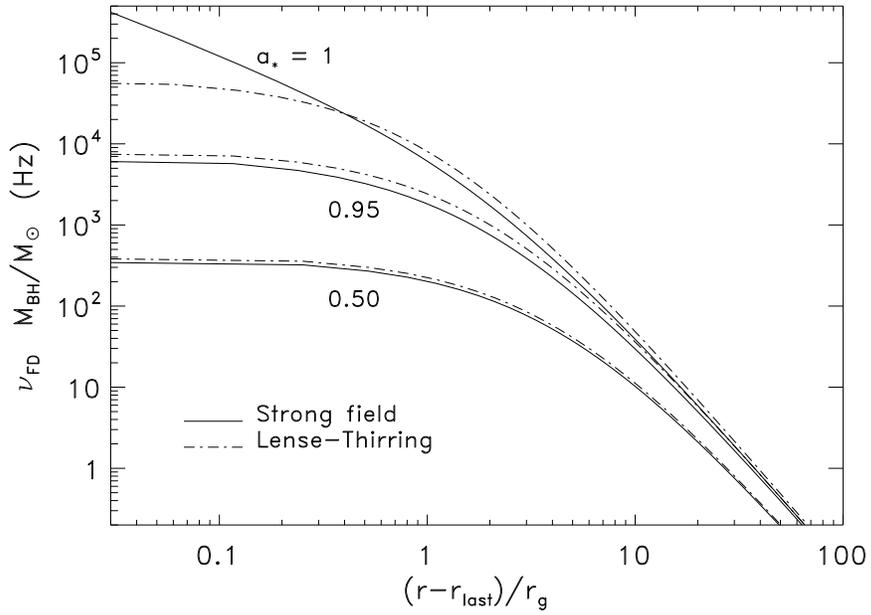,height=9cm}
\caption{Disk precession frequency (multiplied by black hole mass), as a
function of distance from the last stable orbit, for different BH angular
momentum (assuming $Q=0.01$; see text). The weak-field limits are also shown 
in dash-dotted line for comparison. }
\end{figure}

\clearpage
\begin{figure}
\psfig{figure=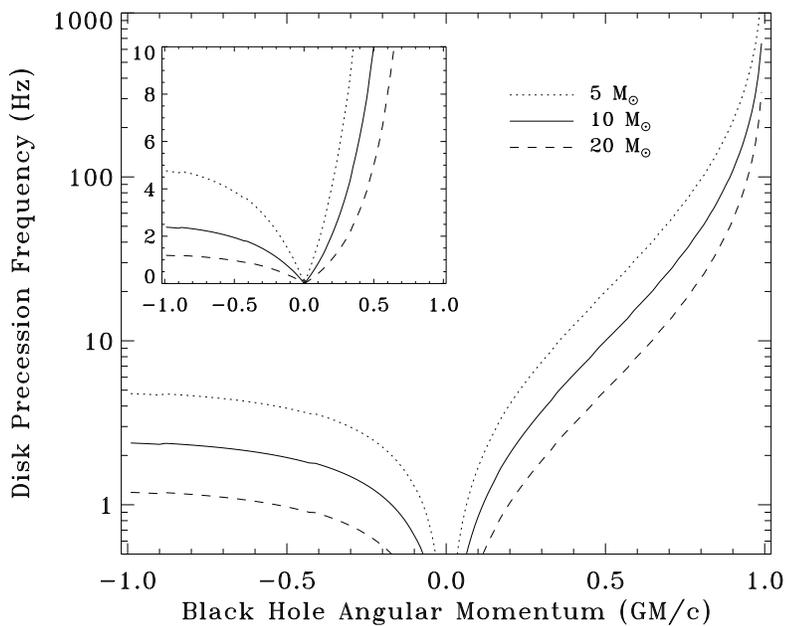,height=9cm}
\caption{Disk precession frequency at where the disk emission peaks 
($r_{\rm peak}$), as a function of the dimensionless specific angular 
momentum ($a_\ast$) of Kerr black holes (assuming $Q=0.01$). }
\end{figure}


\end{document}